\documentclass[twocolumn,showpacs,prb,amsmath,amssymb,superscriptaddress,floatfix]{revtex4}
\usepackage{graphicx,bm,color}

\begin{document}

\title{Evidence for temperature dependent spin-diffusion as a mechanism of intrinsic flux noise in SQUIDs}
\author{T. Lanting}
\affiliation{D-Wave Systems Inc., 3033 Beta Avenue, Burnaby, BC V5G 4M9, Canada}
\author{M.H. Amin}
\affiliation{D-Wave Systems Inc., 3033 Beta Avenue, Burnaby, BC V5G 4M9, Canada}
\affiliation{Department of Physics, Simon Fraser University, Burnaby, British Columbia, Canada V5A 1S6}
\author{A.J. Berkley}
\affiliation{D-Wave Systems Inc., 3033 Beta Avenue, Burnaby, BC V5G 4M9, Canada}
\author{C. Rich}
\affiliation{D-Wave Systems Inc., 3033 Beta Avenue, Burnaby, BC V5G 4M9, Canada}
\author{S.-F. Chen}
\author{S. LaForest}
\affiliation{Department of
  Physics and Astronomy, University of Victoria, Victoria, B.C., V8W
  2Y2, Canada}
\author{Rog\'{e}rio de Sousa} \email[]{rdesousa@uvic.ca} 
\affiliation{Department of
  Physics and Astronomy, University of Victoria, Victoria, B.C., V8W
  2Y2, Canada}

\date{\today}

\begin{abstract}
  The intrinsic flux noise observed in superconducting quantum
  interference devices (SQUIDs) is thought to be due to the
  fluctuation of electron spin impurities, but the frequency and
  temperature dependence observed in experiments do not agree with the usual $1/f$ models. We present theoretical calculations and experimental measurements of flux noise in rf-SQUID flux qubits that show how these observations, and previous reported measurements, can be interpreted in terms of a spin-diffusion constant that increases with temperature. We fit measurements of flux noise in sixteen devices, taken in the $20-80$~mK temperature range, to the spin-diffusion model. This allowed us to extract the spin-diffusion constant and its temperature dependence, suggesting that the spin system is close to a spin-glass phase transition.
\end{abstract}


\maketitle

\section{Introduction}

Low frequency fluctuations of magnetic flux are a dominant noise
source in a wide range of superconducting circuits, including
SQUID-based magnetometers and rf-SQUID flux qubits used as the
building blocks for quantum computing architectures.\cite{clarke08}
In the case of qubits, this magnetic flux noise places fundamental
limits on the performance and scalability of such architectures. Low
frequency flux noise is widely thought to be due to fluctuations of
magnetic impurities local to the superconductor wiring\cite{koch07,bialczak07,sendelbach08} but the
identity of these impurities and the physical mechanism producing the
observed fluctuations is not known. Understanding the fundamental
origin of flux noise is important not only to aid in its reduction in
superconducting devices, but also may provide insight into the
behavior of disordered ensembles of spins at low temperature. Indeed,
flux noise experiments\cite{sendelbach08} suggested the presence of a
spin-glass phase, motivating further computational studies of
fluctuations in spin-glasses.\cite{chen10}

Some likely candidates for spin impurities include dangling-bonds in
the oxide surrounding the superconducting wire,\cite{desousa07} or
disorder-induced localized states at the superconductor-insulator
interface.\cite{choi09} The key point is that the spins must be
located close to the superconducting wire for their flux to be
significant. Any mechanism that produces spin dynamics will contribute
to flux noise.  Since spin-lattice relaxation is suppressed at low
temperatures,\cite{desousa07} it was proposed that the RKKY
interaction between spins is responsible for spin-diffusion and flux
noise;\cite{faoro08} however their theory of spin-diffusion predicted
$1/f$ noise independent of temperature, thus it does not explain the
frequency and temperature dependence of flux noise observed in SQUIDs.\cite{wellstood87,anton13}

Anton {\it et al.}\cite{anton13} recently presented a comprehensive experiment that seemed to contradict the spin-diffusion interpretation; they
measured flux noise above $T=100$~mK and proposed a power law fit of
the form $A^{2}/f^{\alpha}$, showing that the temperature dependence
of $A$ and $\alpha$ leads to the presence of a pivot frequency, below
(above) which the noise decreases (increases) with increasing
temperature.
 
In this article, we propose a theory of spin-diffusion in SQUIDs that
explains these observations. We predict the presence of a crossing
band supported by Ref.~\onlinecite{anton13} and present additional experimental
measurements of flux noise in superconducting flux qubits in a
temperature range between 20 and 80 mK that confirm this prediction.
The model allows us to use these measurements to estimate the
spin-diffusion constant and explore its dependence on temperature.

\section{Flux produced by spins near SQUID wiring}
\label{sec:fluxcalc}

We assume an
ensemble of spin-$1/2$ impurities distributed nearby the SQUID wire;
each spin is located at $\bm{R}_i$, and described by the dimensionless
spin operator $\bm{s}_i$.  The SQUID detects a total flux of
\begin{equation}
\Phi(t)=\sum_i \bm{F}(\bm{R}_i) \cdot \bm{s}_i(t), 
\end{equation}
with a flux
vector $\bm{F}(\bm{R}_{i})$ representing the dependence of the flux on
different spin orientations.

We can find an explicit expression for $\bm{F}(\bm{R}_i)$ by noting that
the coupling energy between the spin and the SQUID is ${\cal
  H}_{\rm{s-SQ}}=-\bm{\mu}_i\cdot \bm{B}(\bm{R}_i)$, where
$\bm{\mu}_i=-g\mu_B \bm{s}_i$ is the magnetic moment of the electron
spin, $g\approx 2$ is its $g$-factor, $\mu_B$ is the Bohr
magneton, and
\begin{equation}
\bm{B}(\bm{R}_i)=\frac{\mu_0}{4\pi}\int d^3 r \frac{\left(\bm{r}-\bm{R}_i\right)\times\bm{J}_{\rm{SQUID}}(\bm{r})}
{\left|\bm{r}-\bm{R}_i\right|^{3}}
\label{eq:br}
\end{equation}
is the magnetic field produced by the SQUID's current density
$\bm{J}_{\rm{SQUID}}(\bm{r})$. The flux-inductance theorem (proven in
Section 5.17 of Ref.~\onlinecite{jackson99}) implies that the flux produced
by the spin must relate to the coupling energy according to
$\Phi_i={\cal H}_{\rm{s-SQ}}/I_{\rm{SQUID}}$, where $I_{\rm{SQUID}}$
is the total current flowing through the SQUID's loop. From this we
get an explicit expression for $\bm{F}(\bm{R}_i)$:
\begin{equation}
\bm{F}(\bm{R}_i)=\frac{g\mu_B\mu_0}{4\pi} \int d^3 r \frac{(\bm{r}-\bm{R}_i)
\times \bm{J}_{\rm{SQUID}}(\bm{r})}{\left|\bm{r}-\bm{R}_i\right|^{3}I_{\rm{SQUID}}}.
\label{fvec}
\end{equation}

We performed explicit numerical calculations of Eq.~(\ref{fvec}) for
the particular geometry of the devices we tested. We provide details on the geometry and the numerical calculations in Appendix~\ref{sec:appendix-geometry}.

\section{Model of flux noise due to spin-diffusion}

The thermal equilibrium flux noise
\begin{equation}
\tilde{S}_{\Phi}(f)=\int_{-\infty}^{\infty} dt \;\textrm{e}^{2\pi i f t}\langle \Phi(t)\Phi(0)\rangle,
\label{stildephi}
\end{equation}
can be computed in the so called spin-diffusion regime, where
only long wavelength fluctuations of the spin system are taken into
account. This is done by considering a coarse-grained
impurity spin field, $\bm{M}(\bm{r},t)=\sum_{i}\bm{s}_i
\delta(\bm{r}-\bm{R}_i)$, with the flux written as
\begin{eqnarray} 
\Phi(t)&=&\int d^2 r \bm{F}(\bm{r})\cdot \bm{M}(\bm{r},t) \nonumber\\
&=& \frac{1}{(2\pi)^{2}}\int d^{2}k \int df \textrm{e}^{-2\pi i f t} \bm{\tilde{F}}^{*}(\bm{k})\cdot \bm{\tilde{M}}(\bm{k},f), 
\label{fs}
\end{eqnarray}
with $\bm{\tilde{F}}(\bm{k})=\int d^{2}r\;\textrm{e}^{-i\bm{k}\cdot\bm{r}}\bm{F}(\bm{r})$ and $\bm{\tilde{M}}(\bm{k},f)=\int d^{2}r\int dt\;\textrm{e}^{-i(\bm{k}\cdot\bm{r}-2\pi f t)}\bm{M}(\bm{r},t)$.

We consider a model Hamiltonian of exchange coupled spins, 
\begin{equation}
{\cal  H}_{\rm{s-s}}=\sum_{i<j}J_{ij}\bm{s}_i\cdot \bm{s}_j. 
\end{equation}
In the paramagnetic phase (where
$\langle \bm{s}_i\rangle=0$) the spin field will satisfy the 
following Langevin equation,\cite{chaikin00}
\begin{equation}
\frac{\partial \bm{M}}{\partial t}=D \nabla^{2}\bm{M}+ \bm{\zeta},
\label{sdeqn}
\end{equation}
where $D$ is a diffusion constant (possibly temperature dependent\cite{bennet65}),
and $\bm{\zeta}$ is a random force that drives the spins into thermal
equilibrium with themselves. 
This implies the following correlation for
the random force,\cite{chaikin00}
\begin{equation}
\left\langle \zeta_\alpha(\bm{r},t)\zeta_\beta(\bm{r}',t')\right\rangle = 
\frac{\sigma D}{2}\frac{\chi}{\chi_0}
\nabla^{2}\delta(\bm{r}-\bm{r'})\delta(t-t')\delta_{\alpha\beta},
\label{zetacor}
\end{equation}
where $\sigma$ is the area density for spins, $\chi$ is the interacting spin susceptibility (defined as $\chi=\partial\langle M\rangle/\partial B$ at $B=0$), and $\chi_0=-g\mu_B \sigma/(4k_BT)$ is the free spin (Curie) susceptibility.  
This random force correlator
is chosen so that the fluctuation-dissipation theorem is satisfied
(thus leading to the expected thermal equilibrium state at long
times). We emphasize that we assume the spins are decoupled from their
lattice, so that the total magnetization $\sum_i \bm{s}_i$ is
conserved [the presence of $\nabla^{2}$ in  Eq.~(\ref{zetacor}) ensures this conservation law].

Writing Eqs.~(\ref{stildephi})--(\ref{zetacor}) in Fourier space and evaluating the spin-spin correlation function leads to a convenient expression relating SQUID geometry to flux noise:
\begin{equation}
\tilde{S}_{\Phi}(f)= \frac{\sigma}{2(2\pi)^{4}}\frac{\chi}{\chi_0}\int d^2 k |\tilde{\bm{F}}(\bm{k})|^{2}\frac{Dk^2}{f^{2}+(Dk^2/2\pi)^{2}},
\label{stildefinal}
\end{equation}
with the function $|\tilde{\bm{F}}(\bm{k})|^{2}$ playing the role of a ``form factor'' for flux noise. As a check, note that $\int df
\tilde{S}_{\Phi}(f)=\frac{\sigma}{4}\frac{\chi}{\chi_0}\int d^2 r |\bm{F}(\bm{r})|^{2}$,
which at high $T$ (when $\chi\rightarrow \chi_0$) is equal to the expected $\langle \Phi^2\rangle= \sum_{i,j}\sum_{\alpha,\beta}F_{\alpha}(\bm{R}_i)F_{\beta}(\bm{R}_j)\langle s_{i\alpha}s_{j\beta}\rangle=\frac{1}{4}\sum_i |\bm{F}(\bm{R}_{i})|^{2}$. 

Our Eq.~(\ref{stildefinal}) goes beyond the constrained 1d model of Ref.~\onlinecite{faoro08}, allowing spin-diffusion across the whole 2d device
area (this feature of our model can lead to non-zero
  flux noise correlation between two different devices for certain
  geometries and could explain the measurements reported in Ref.~\onlinecite{yoshihara10}). We account for SQUID geometry by introducing
the form factor $|\tilde{\bm{F}}(\bm{k})|^{2}$, that modulates the
weight of each diffusion mode with characteristic frequency
$Dk^{2}/(2\pi)$.  It bears an interesting analogy with optics, since
$|\tilde{\bm{F}}(\bm{k})|^{2}$ is identical to the Fraunhofer
diffraction pattern of an aperture in the shape of the SQUID's wire.
Note how this is quite distinct from the usual model of $1/f$ noise in
electronic systems, where each fluctuation mode is instead weighted by
a probability distribution related to material disorder.\cite{weissman88}

\section{Frequency and temperature dependence of flux noise} 

We consider a rf-SQUID in the shape of a rectangular washer, 
with external sides $L_{\parallel}$ and $L_{\perp}$. The washer is made of thin film wires of thickness $b$ and lateral width $W$, with $W\gg b$ (See Fig.~\ref{fig:squid}). 

Substituting $\bm{k}'=W\bm{k}$, we can rewrite Eq.~(\ref{stildefinal}) as
\begin{equation}
\tilde{S}_{\Phi}(f)=\frac{\sigma\chi}{8\pi^{2}\chi_0 D}\int d^{2}k'|\tilde{\bm{F}}(\bm{k'})|^{2} \frac{k'^{2}}{(f/f_c)^{2}+k'^{4}},
\label{stilderenormalized}
\end{equation}
where we defined the characteristic frequency $f_c= D/(2\pi W^{2})$, with $1/f_c$ describing the time scale for non-equilibrium spin polarization to diffuse across the SQUID's wire width $W$. 
From Eq.~(\ref{stilderenormalized}) we see that increasing (decreasing) $D$ shifts the spectrum to higher (lower) frequencies, with its area $\langle\Phi^{2}\rangle$ remaining constant. We verified that the total noise power is independent of $D$ and scales as $\langle\Phi^{2}\rangle\propto (L_{\parallel}+L_{\perp}-2W)/W$, in agreement with Ref.~\onlinecite{bialczak07}.

\begin{figure}
\includegraphics[width=3.4in]{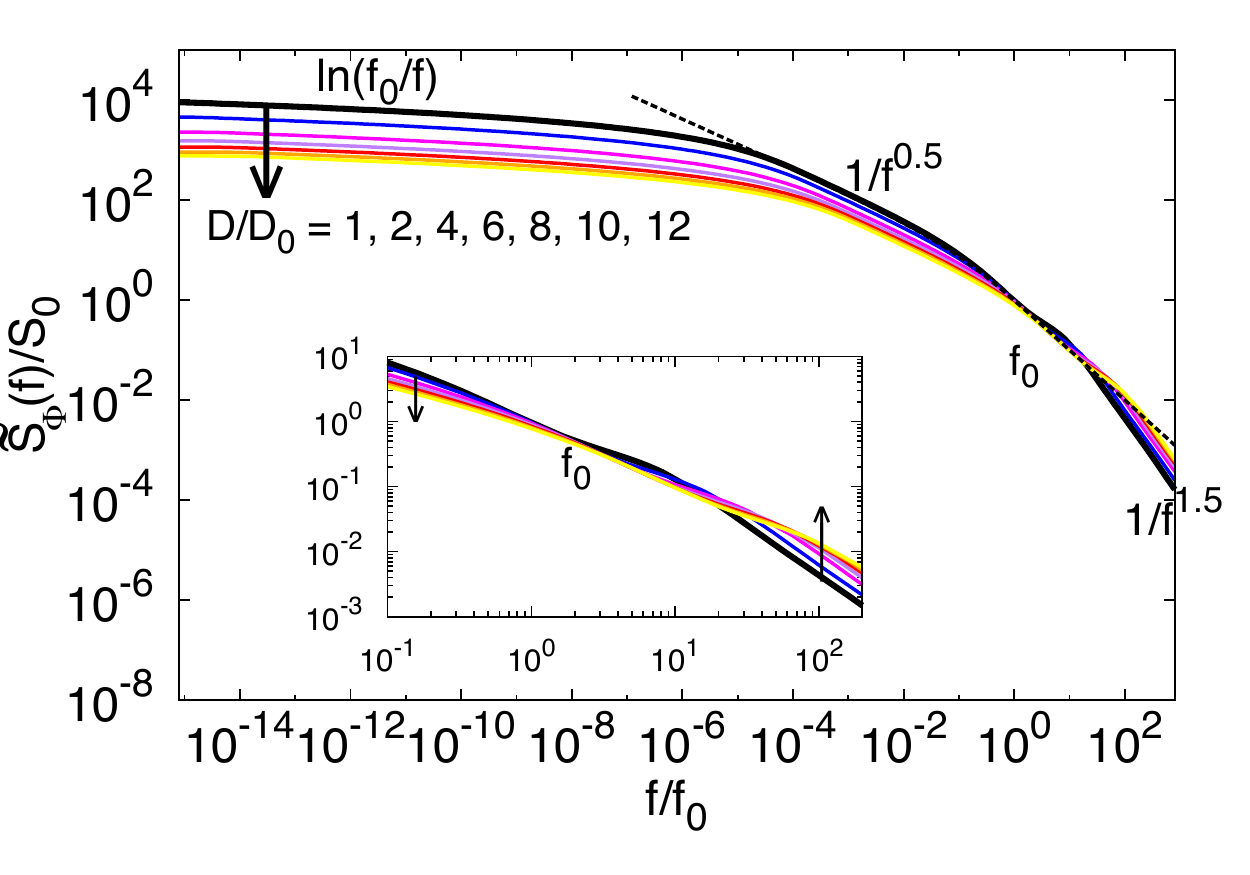}
\caption{(color online). Theoretical calculation of flux noise due to spin-diffusion in a rectangular SQUID of sides $L_{\parallel}$ and $L_{\perp}$, with wire width $W$ (we used $L_{\parallel}/L_{\perp}=350/3$ and $L_{\perp}/W=3$ to compare to the experiments below). The solid curves show $\tilde S_{\Phi}(f)$ for a range of $D$. The axes are normalized to $f_0$ and $S_0 \equiv \tilde S_{\Phi}(f_0)$, where $f_0$ is chosen to be the point where the curves appear to cross. We also normalize $D$ by the quantity $D_0 = 2\pi W^2 f_0$. The dashed line shows the fitting function used in Eq.~(\ref{analytic}). Inset: A zoom into the crossing band region at $f_{0}$. The arrows show the behavior for increasing $D$. At frequencies lower than $f_{0}$, the noise spectral density \emph{decreases} with increasing $D$; at frequencies higher than $f_{0}$ the opposite behaviour takes place.}
\label{figD_Tdepend}
\end{figure}

In Figure~\ref{figD_Tdepend} we show the noise power spectral density calculated for our SQUID geometry using Eq.~(\ref{stilderenormalized}) for a range of $D$. The axes are normalized to $f_0$ and $S_0\equiv \tilde S(f_0)$, where $f_0$ is chosen to be the point where the curves appear to cross. We also normalize $D$ by the quantity $D_0 = 2\pi W^2 f_0$ (note that $f_0$ coincides with $f_c$ when $D=D_0$). 

At low frequencies ($f< 10^{-5} f_{0}$), the noise scales logarithmically as $\ln{(f_{0}/f)}$, flattening out due to the finite size of the SQUID.  At intermediate frequencies ($f=10^{-4}-10^{-1} f_{0}$), it varies as $1/f^{0.5}$ over three decades of frequency; at higher frequencies ($f> 10 f_{0}$), the noise is cut-off as $1/f^{1.5}$.

The low frequency limit of Eq.~(\ref{stildefinal}) depends crucially
on dimensionality. In 1d, low frequency noise diverges as $1/f^{0.5}$,\cite{macfarlane50,scofield85} 
while in 2d it diverges logarithmically as shown in Fig.~\ref{figD_Tdepend}; in 3d the
noise flats out as a constant. On the other hand, the high frequency
limit of Eq.~(\ref{stildefinal}) is $1/f^{1.5}$ for all dimensions.
Notably, in all cases the $1/f$ behavior obtained in a previous model\cite{faoro08} appears only in a narrow frequency band.

The diffusion constant may be temperature dependent.  This gives a possible mechanism for the temperature dependence of flux noise observed in several experiments since Ref.~\onlinecite{wellstood87}. Figure~\ref{figD_Tdepend} shows plots of the spin-diffusion
noise for different values of the diffusion constant $D/D_0=1$--$12$. 
In the range of frequencies $f_c=D/(2\pi W^{2})$ the different curves approach each other and a crossing band occurs. The inset of Fig.~\ref{figD_Tdepend} displays the same calculation focusing on the frequency band where this crossing occurs. $\tilde{S}_{\Phi}(f)$ decreases with increasing $D$ to the left of the crossing band and increases with increasing $D$ in the region to the right of the crossing band.

\section{Measurements of flux noise} 

\begin{figure}
\includegraphics[trim=2cm 6.0cm 2cm 7.0cm,clip,width=8cm]{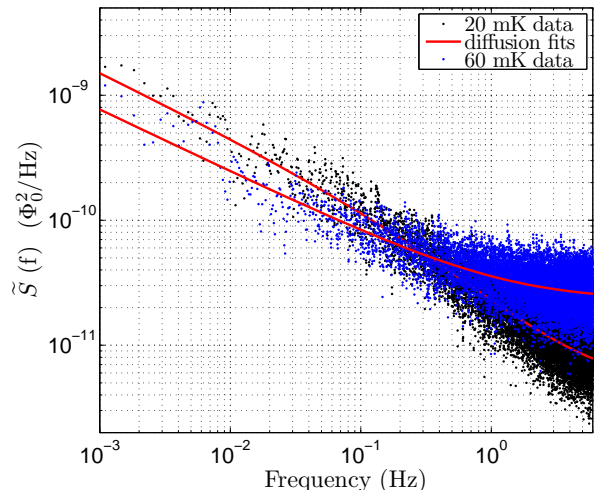}
\caption{\label{raw-psd} (color online). Typical measurements of low frequency noise. We plot $\tilde{S}_{\Phi}(f)$ vs $f$ for one of the qubits ($q2$) at $T=20$  and $60$~mK. The solid curves show fits to Eq.~(\ref{psd}). We observe a crossing at $0.1-1$ Hz across the measured temperature range, roughly consistent with Ref.~\onlinecite{anton13}.}
\end{figure}

To test the spin-diffusion model, we performed measurements of flux noise on sixteen compound Josephson junction rf-SQUID flux
qubits\cite{harris10} with identical geometries, shown in Fig.~\ref{fig:squid}. The devices were fabricated with a process comprising a Nb/AlOx/Nb trilayer, planarized SiO$_2$ dielectric layers and Nb wiring layers. The rf-SQUID wires have lateral width $W = 1$~$\mu$m, with $L_{\parallel} = 350$~$\mu$m and $L_{\perp} = 3$~$\mu$m, and are separated by $0.2$~$\mu$m from a Nb ground plane.
The sample was mounted to the mixing chamber of a
dilution refrigerator with its temperature regulated at set points
between $20$ and $80$ mK. We shielded the sample from external
magnetic flux with a superconducting Al shield.

\begin{figure}
\includegraphics[trim=2cm 6cm 1cm 6.5cm,clip,width=9cm]{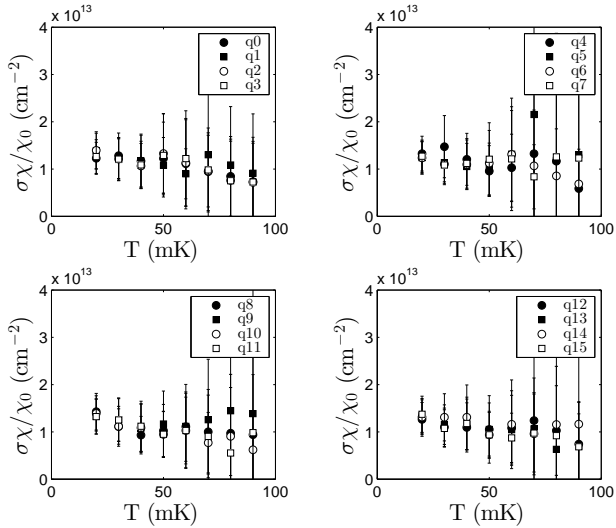}
\caption{Fit values of $\sigma\chi/\chi_0$ versus temperature for sixteen devices. $\sigma\chi/\chi_0$ was obtained by fitting the measured power spectral density to Eq.~(\ref{psd}). Within the experimental error bars, $\sigma \chi/\chi_0$ seems to be independent of temperature and device. For $\chi = \chi_0$, this suggests the spin areal density, $\sigma$, is constant.
\label{S0f0-vs-T}}
\end{figure}

We measured flux noise using a method
described in detail elsewhere.\cite{lanting-2009} We directly
measured the noise power spectral density $\rm{PSD}(f)$ between
$f=1$~mHz and $f=20$~Hz for sixteen devices for temperatures between
$20$~mK and $80$~mK. Our measurements have a white noise contribution
(i.e. frequency independent) $w_n\propto T^{2}$ due to the 
statistical uncertainty of the noise detection method, as well as the low frequency noise from the devices (Appendix~\ref{sec:appendix-measurements} describes the origin of the white noise background $w_n$ and shows that it scales as $T^2$). We extract $\tilde{S}_{\Phi}(f)$ by fitting our data to
\begin{equation}
{\rm PSD}(f)= \tilde{S}_{\rm analytic}(f) + w_n,
\label{psd}
\end{equation}
where $w_n$ is this white noise contribution 
and $\tilde{S}_{\rm analytic}(f)$ is an analytic approximation to Eq.~(\ref{stilderenormalized}) given by
\begin{equation}
\tilde{S}_{\rm analytic}(f) = \left(\frac{\sigma \chi}{\chi_0}\right)\frac{F_0^2L_{\perp}^2}{f}\left(1-e^{-\sqrt{\xi 2\pi W^2 f/D}}\right).
\label{analytic}
\end{equation}
The parameter $\xi=17.6$ is a numerical fit to Eq.~(\ref{stilderenormalized}) 
calculated for the rf-SQUID geometry discussed above, and $F_0=4 ~{\rm n}\Phi_0$ is the value of the modulus of the flux vector for surface spins obtained in Appendix~\ref{sec:appendix-geometry}. Equation~(\ref{analytic}) is shown as a dashed line in 
Fig.~\ref{figD_Tdepend}, where it is seen to fit the noise spectrum over a limited range of frequencies ($10^{-4} f_{0}<f<10 f_{0}$), that turns out to be the relevant range measured in our experiments.
We fit our data to Eq.~(\ref{psd}) in order to extract the two free parameters, $D$ and $\left(\frac{\sigma \chi}{\chi_0}\right)$ independently. See Fig.~\ref{raw-psd} for typical data at two different temperatures and fits to Eq.~(\ref{psd}). Note that the crossing band occurs at $f_c = 0.1-1$ Hz.

\begin{figure}
\includegraphics[trim=2cm 6cm 1cm 6.5cm,clip,width=9cm]{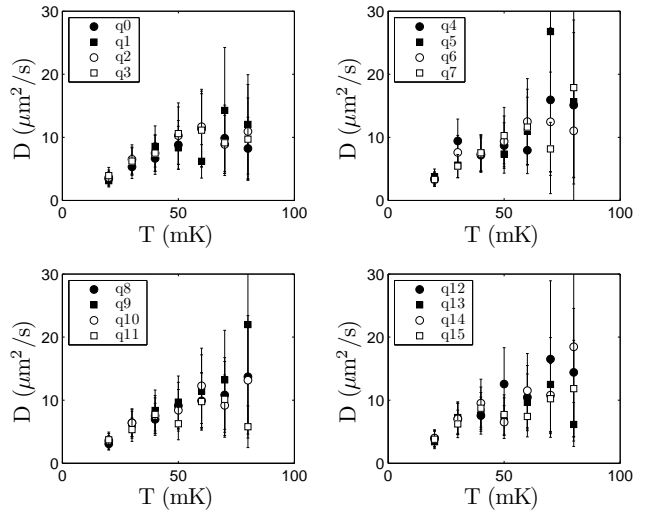}
\caption{\label{D} Fit value of the spin-diffusion constant $D$ versus temperature for sixteen devices. $D$ was obtained by fitting the measured power spectral density to Eq.~(\ref{psd}). In the low temperature range, $D$ shows a clear trend to increase with temperature. In the higher temperature range the increasing contribution of the white noise background $w_n$ increases the error bars making this trend less clear.}
\label{D-vs-T}
\end{figure}

Figure~\ref{S0f0-vs-T} shows the fit quantity $\sigma\chi/\chi_0$ as a function of temperature for sixteen devices.
Within the experimental error bars we see that $\sigma\chi/\chi_0 = 10^{-13}{\rm cm}^{-2}$ is the same constant for all devices, independent of temperature. Plots of $\sqrt{w_n}$ as
  a function of $T$ demonstrate that the qubit
  and refrigerator thermometry were in agreement even at
  the lowest temperatures (See Fig.~\ref{fig:sqrtwn} in the Appendix below). Our theory predicts $\sigma \chi/\chi_0$ independent of $D$.  \emph{Hence Fig.~\ref{S0f0-vs-T} shows that $\sigma\chi/\chi_0$ is roughly independent of $T$, and that any $T$-dependence must originate from $D(T)$}. Assuming $\chi\approx \chi_0$ (paramagnetic phase), we get $\sigma=1\times 10^{13}/{\rm
    cm}^{2}$ for the area density of spins covering the wire (top plus
  bottom).

Fig.~\ref{D-vs-T} shows the fit parameter $D$, the spin diffusion constant, as a function of $T$ for all sixteen devices. The fit values are in the range of $3-30$~$\mu$m$^{2}/$s and show a trend of increasing with temperature.  We only fit data up to
$T=80$~mK. Above this temperature the white noise contribution
$w_n\propto T^{2}$ begins to dominate across the fit
bandwidth and we cannot reliably separate the two terms in
Eq.~(\ref{psd}) and thus extract the intrinsic flux noise
$\tilde{S}_{\Phi}(f)$. At higher temperatures the increasing $w_n$
increases the uncertainty on the fit parameters $D$ and $\sigma\chi/\chi_0$, as
shown by the increasing error bars in
Figs.~\ref{S0f0-vs-T}~and~\ref{D-vs-T}.

\begin{figure}
\includegraphics[trim=2cm 6cm 1cm 6.5cm,clip,width=9cm]{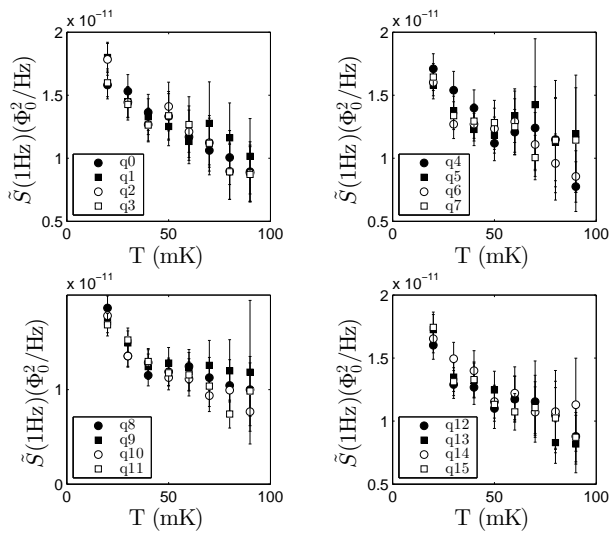}
\caption{$\tilde{S}_{\Phi}(1{\rm ~Hz})$ versus $T$ for all sixteen devices. We used the fit parameters shown in Figs.~\ref{S0f0-vs-T} and \ref{D-vs-T} and Eq.~(\ref{analytic}) to calculate $\tilde{S}_{\Phi}(1~{\rm Hz})$. Note that $\tilde{S}_{\Phi}(1~{\rm Hz})$ decreases with increasing $T$.\label{S0-vs-T}}
\end{figure}

For comparison to measurements by other groups, we show measurements of $\tilde{S}_{\Phi}(1~{\rm Hz})$ versus temperature for all devices in Fig.~\ref{S0-vs-T}. Here we see that there is a clear
temperature dependence, with the PSD decreasing from $1.8\times
10^{-11}$~$\Phi_{0}^{2}/$Hz to $1\times 10^{-11}$~$\Phi_{0}^{2}/$Hz over this range of temperature. 

\section{Interpretation of the experiment: Proximity to a phase transition}

It is known that $D$ is temperature dependent when the spin system is
close to a phase transition.\cite{bennet65,halperin67,sompolinsky81}
Assuming $\sigma$ is constant,
Fig.~\ref{S0f0-vs-T} suggests that the susceptibility is following the
Curie $1/T$ law with no additional temperature dependence. This is
consistent with the behavior just above a spin-glass critical
temperature $T_c$, whose $\chi$ deviates from $\chi_0$ by a kink at
$T=T_c$. The prediction of the theory of dynamical critical phenomena
is that $D\propto |T-T_c|^{1}$ close to the spin-glass transition.\cite{sompolinsky81} Thus the trend in Fig.~\ref{D-vs-T} is
consistent with $T>T_c$ for a spin-glass phase transition.

The scenario of proximity to a phase transition would also be
consistent with previous experiments. It is possible that other
fabrication methods will yield devices with extremely low $T_c$, so
that $T\gg T_c$; in this case $D$ will be independent of $T$ and our
model leads to flux noise that does not change with temperature, as
observed in another recent experiment.\cite{yan12}  In other samples
$T_c$ may be higher, leading to additional $T$-dependence in the
$T\approx T_c$ regime [$\chi\neq\chi_0$ in Eq.~(\ref{stildefinal})],
where spin-clusters are present as was claimed in Ref.~\onlinecite{anton13}.

\section{Conclusions}

In conclusion, our theory of SQUID flux noise due to the
spin-diffusion of interacting spins shows how the frequency and
temperature dependence of flux noise is influenced by SQUID geometry
and a $T$-dependent spin-diffusion constant, giving rise to the
presence of a \emph{crossing band} of frequencies, like the one
observed in a previous experiment.\cite{anton13}  We presented
experimental data in the low temperature range ($T=20-80$~mK), showing
that the theory can explain the experiments provided that we assume
that the spin-diffusion constant increases with temperature. 
A temperature dependent spin diffusion constant suggests that the spin system is close to a spin-glass phase transition, but more experiments are needed to confirm this assertion. 

We thank C. Dasgupta, P. Kovtun, A.Yu. Smirnov, and N.M. Zimmerman for useful discussions. Our research was supported by the NSERC Engage program. 

\appendix

\section{Flux from a spin near SQUID wiring}
\label{sec:appendix-geometry}

\begin{figure}
\includegraphics[width=0.49\textwidth]{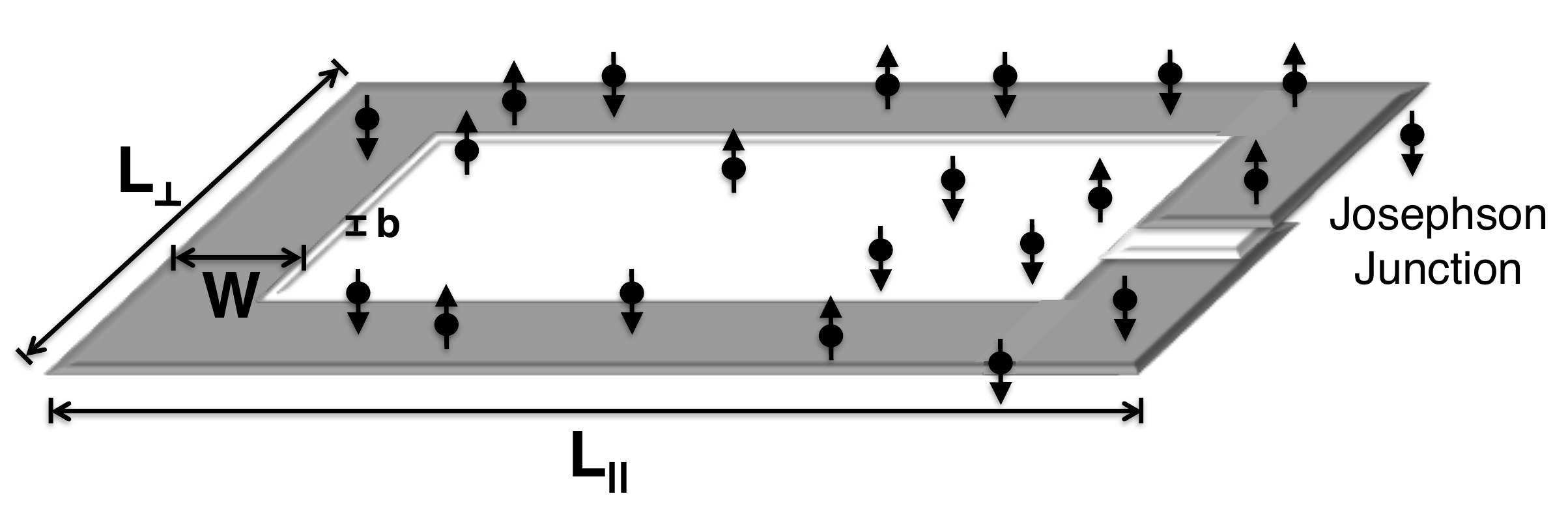}
\caption{An illustration of the geometry of the rf-SQUID flux qubit wiring used in our theoretical calculations and experimental measurements. Our SQUIDs had wire width $W=1$~$\mu$m, wire thickness $b=0.2$~$\mu$m, and lateral dimensions $L_{\parallel}=350~\mu$m and
$L_{\perp}=3~\mu$m. For simplicity we include a single Josephson junction in the illustration. The measurements reported herein were taken with compound-compound Josephson junction devices described in detail elsewhere~\cite{harris10}.}
\label{fig:squid}
\end{figure}

The devices tested were rf-SQUID flux qubits with superconducting wiring (Nb)
in the shape of a rectangular washer, with external sides $L_{\parallel}=350~\mu$m and
$L_{\perp}=3~\mu$m. The washer is made of thin film wires of thickness $b=0.2$~$\mu$m and 
lateral width $W=1$~$\mu$m (See Fig.~\ref{fig:squid}). 

We used the analytical expression for the current
density flowing through a thin film superconductor (with wire width
$W$ much larger than film thickness $b$, just like our SQUID wires). 
We used the full interpolated expression described in Eqs.~(1)--(4) of
Ref.~\onlinecite{vanduzer99} for $\bm{J}_{\rm{SQUID}}(\bm{r})$.  The
results for $|\bm{F}(\bm{R})|$ as a function of spin position $\bm{R}$
for wire width $W=1$~$\mu$m and wire thickness $b=0.2$~$\mu$m is shown
in Fig.~\ref{fig:f} below.  As a check of the reliability of our
results, we performed computations for wires and spin distances
similar to the ones considered in Ref.~\onlinecite{koch07}, confirming
that our calculations are in good agreement with previous
calculations performed with the software package FastHenry.

\begin{figure}
\centering
\includegraphics[width=0.49\textwidth]{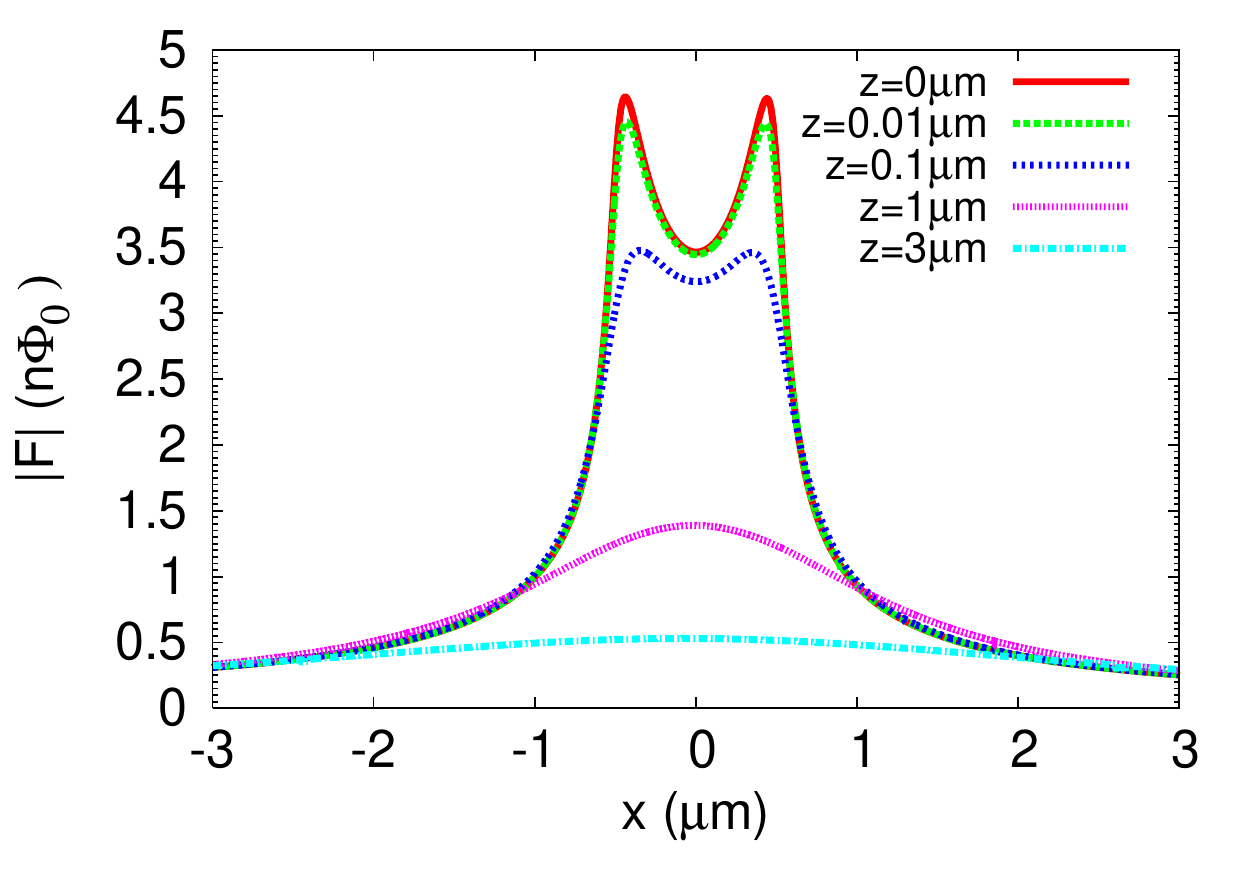}
\caption{(color online). Calculations of the modulus of the flux vector $\bm{F}(\bm{r})$ as a function of spin position $\bm{r}$ for the rf-SQUID flux qubit geometry of Fig.~\ref{fig:squid}.  Here $x$ is the spin's coordinate along the SC wire width (the wire edges are at $x=\pm W/2$), and  $z$
is the spin coordinate perpendicular to the wire ($z=0$ is at the wire surface). 
Note that at the wire surface, $|\bm{F}|$ is within 20\% of $F_0 = 4$~n$\Phi_0$.}
\label{fig:f}
\end{figure}

The qubit wiring is separated by $0.2\ \mu$m from a Nb superconducting ground plane in order to provide magnetic shielding from other flux sources. The  effect of this ground plane is to produce a mirror current distribution in the plane below the SQUID. Because of symmetry, the current distribution is not changed (i.e., the current is still peaked at the edge of the SQUID wire). While the ground plane does affect the value of the flux produced by a spin away from the wire surface (e.g. spins in the midpoint between the SQUID and the ground plane will have $\bm{F}=\bm{0}$), it does not significantly affect the value of flux vector $\bm{F}$ for the spins located at the wire surface ($z=0$ in Fig.~\ref{fig:f}).  This happens because for surface spins, the contribution from the mirror current is insignificant in comparison to the contribution from the actual SQUID wire current. 
Note how in Fig.~\ref{fig:f} the value of $|\bm{F}|$ decreases rapidly as the spin-wire distance increases (In Fig.~\ref{fig:f} only spins within $\sim 0.1$~$\mu$m of the wire surface produce appreciable flux). 

The result that $\bm{F}$ decreases rapidly with increasing spin-wire distance motivates the consideration of a 2d model of surface or interface spins with $\bm{R}_{i}$ in the plane of the wire. 
In the present calculations we took $|\bm{F}(\bm{R}_{i})|\approx F_0$ when $\bm{R}_{i}$ is on the surface of the wire, and $\bm{F}(\bm{R}_{i})=\bm{0}$ elsewhere.
In Fig.~\ref{fig:f} we see that surface spins have $|\bm{F}|=F_0=4\ {\rm n}\Phi_0$ apart from an oscillation of 20\%. 

\section{Flux noise measurement details}
\label{sec:appendix-measurements}

Here we describe the method used to measure flux noise in more detail. The method was
introduced in Ref.~\onlinecite{lanting-2009} as a method of detecting 1/f noise {\em in situ} in superconducting flux qubits. The qubit design is described in detail in Ref.~\onlinecite{harris10}.

\subsection{The CCJJ qubit}

The qubits are compound-compound Josephson-Junction (CCJJ) rf-SQUID flux qubits described in Ref.~\onlinecite{harris10}. Two external flux biases $\Phi_{\rm CCJJ}^x$ and $\Phi_q^x$ allow us to operate the qubit as an effective Ising spin governed by the Hamiltonian:

\begin{equation}
{\cal H}=-\frac{1}{2}\left[\epsilon\sigma_z +\Delta\sigma_x\right],
\end{equation}

where $\epsilon = 2|I_q^p|(\Phi_q^x-\Phi_q^0)$, $|I_q^p(\Phi_{\rm CCJJ}^x)|$ is the expectation value of persistent current and $\Delta(\Phi_{\rm CCJJ}^x)$ is the tunneling amplitude. We anneal the qubit by ramping $\Phi_{\rm CCJJ}$ from $\Phi_0/2$ to $\Phi_0$ in a time $t_a = 10\ \mu$s and then read out its state. When the qubit is in thermal equilibrium with a thermal bath at temperature $T$, the probability of detecting state $\left\vert\uparrow\right\rangle$ is given by:

\begin{equation}
P=\frac{\textrm{e}^{\epsilon/(2k_BT)}}{\textrm{e}^{\epsilon/(2k_BT)}+\textrm{e}^{-\epsilon/(2k_BT)}}
=\frac{1}{2}\left[1+\tanh{\left(\frac{\Phi_q^x-\Phi_q^0}{2\delta}\right)}\right],
\label{eq:p1}
\end{equation}
where $\delta$ is given by 
\begin{equation}
\delta=\frac{k_BT}{2|I_q^{p*}|},
\label{eq:delta}
\end{equation}
and where $|I_q^{p*}|$ is the persistent current of the qubit at the point where dynamics cease and the qubit localizes into $\left\vert\uparrow\right\rangle$ or $\left\vert\downarrow\right\rangle$.

\subsection{Noise Measurement Method}

The noise measurement technique is described in detail in Ref.~\onlinecite{lanting-2009}. In the presence of a low frequency noise signal $\Phi_n(t) \ll \delta$, Eq.~\ref{eq:p1} becomes

\begin{equation}
P=\frac{1}{2}\left[1+\tanh{\left(\frac{\Phi_n(t)+\Phi_q^x-\Phi_q^0}{2\delta}\right)}\right]
\label{eq:p1-noise}
\end{equation}

We first calibrate $\Phi_q^0$ and $\delta$ and then perform measurements of $P(t)$ by setting $\Phi_q^x = \Phi_q^0$ and then annealing and reading out the qubit $n$ times. Whenever the
outcome is $\left\vert\uparrow\right\rangle$, we assign a value $P_i=1$, and when the
outcome is $\left\vert\downarrow\right\rangle$ we assign $P_i=0$. After $n$ anneals we get 
\begin{equation}
P=\frac{1}{n}\sum_{i=1}^{n}P_i.
\label{eq:p1stat}
\end{equation}
We then convert this into a measurement of $\Phi_n$ by using
Eq.~(\ref{eq:p1-noise}). We repeat this
procedure $m$ times (a total of $n\times m$ anneals) in order to get a
sampling of $\Phi_n(t)$ at time intervals $t_j=j\Delta t$,
$j=1,2,\ldots,m$. Here, $\Delta t = n\tau_s$ where $\tau_s$ is the time required to anneal and read out the qubit once. We then take the fast Fourier transform of the
sample and extract the noise as
\begin{equation}
\tilde{S}_{\Phi}(f_k)=\frac{1}{m\Delta t}\left|\tilde{\Phi_n}(f_k)\right|^{2}, 
\end{equation}
at frequencies $f_k=\frac{k}{m\Delta t}$,
$k=-\frac{m}{2}+1,-\frac{m}{2}+2,\ldots,\frac{m}{2}$. The highest
frequency is the Nyquist frequency $f_{\rm{Nyquist}}=\frac{1}{2\Delta t}$. 

\begin{figure}
\includegraphics[trim=2cm 6cm 1cm 6.5cm,clip,width=9cm]{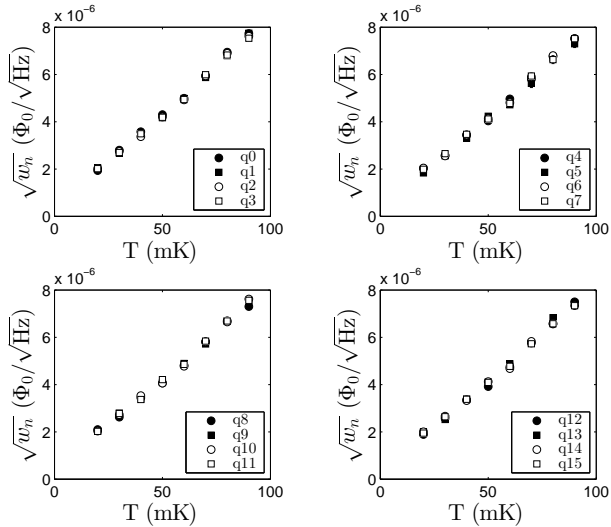}
\caption{Plots of $\sqrt{w_n}$ vs. refrigerator thermometry $T$ for sixteen qubits. The data scale as expected from Eq.~\ref{eq:wn}, direct evidence that the qubit temperature matches the refrigerator temperature. Note in particular that $\sqrt{w_n}$ does
  not saturate at low $T$, demonstrating that the qubit temperature and
  refrigerator thermometry are in agreement even at the lowest
  temperatures.\label{fig:sqrtwn}}
\end{figure}

The statistical limit of this noise detection procedure is related to
the variance in the measurement of $P$ using Eq.~(\ref{eq:p1stat}).
Denoting $p\equiv \langle P\rangle$ for the probability of measuring
outcome $P_i=1$:
\begin{equation}
\sigma_{P}^{2}=\langle P^{2}\rangle-\langle P\rangle^{2}=\left(\frac{p(1-p)}{n}+p^{2}\right)-p^{2}=\frac{p(1-p)}{n}\approx \frac{1}{4n},
\end{equation}
the last approximation when $p\approx 1/2$. This variance will lead to
a white noise background $w_n$ for the measurement of
$\tilde{S}_{\Phi}(f)$. To compute $w_n$ note that
\begin{equation}
\langle \Phi_{n}^{2}\rangle = \left(4\delta\right)^{2}\sigma_{P}^{2}=\frac{4\delta^{2}}{n}
=\int_{-f_{\rm{Nyquist}}}^{f_{\rm{Nyquist}}}w_n=2f_{\rm{Nyquist}}w_n.
\end{equation}
We can then obtain an expression for $w_n$:
\begin{equation}
w_n=4\tau_s \delta^2 \propto T^2.
\label{eq:wn}
\end{equation}
The result that $w_n$ scales with temperature squared gives a useful
means of comparing qubit temperature and refrigerator temperature. Figure~\ref{fig:sqrtwn} shows a plot of $\sqrt{w_n}$ as a
function of $T$. The linear relation evident in Figure~\ref{fig:sqrtwn} is direct evidence that the qubit is in thermal equilibrium with its environment at all temperatures investigated.

\end{document}